\documentclass{aplopt}   
\usepackage{psfig}
\usepackage{epsfig}
\usepackage{graphicx}
\usepackage{natbib}   
\bibpunct{(}{)}{;}{a}{}{,} 
\def \t2pi {$t_{2\pi}$}
\def \mt2pi {t_{2\pi}}    
\def \muas {$\mu''$}

\begin{document}

\titlerunning{ X-ray and Gamma-ray Fresnel lenses}
\authorrunning{G. K. Skinner}

\title{Design and imaging performance of achromatic diffractive/refractive X-ray and Gamma-ray Fresnel lenses}

\author{Gerald K. Skinner}

\institute
{Centre d'Etude Spatiale des Rayonnements, 9, avenue du Colonel Roche, 
31028, Toulouse, France}

\abstract {
 {Achromatic combinations of a diffractive Phase Fresnel Lens and a refractive correcting element have been proposed for X-ray and gamma-ray astronomy and for microlithography, but considerations of absorption often dictate that the refractive component
be given a stepped profile, resulting in a double Fresnel lens.
The imaging performance of corrected Fresnel lenses, with and
without `stepping' is investigated and the trade-off between
resolution and useful bandwidth in different circumstances is
discussed. Provided the focal ratio is large, correction lenses
made of low atomic number materials can be used with X-rays in the
range approximately 10--100 keV without stepping. The use of
stepping extends the possibility  of correction to higher aperture
systems, to energies as low as a few kilo electron volts and  to
gamma-rays of $\sim$ mega electron volt energy.} }

\maketitle 

\section{Introduction}

Zone plates and Fresnel lenses are widely used at X-ray
wavelengths for microscopy and microlithography (e.g. references
 \cite {eg1,eg2,eg3}
 )
and have been proposed for use in X-ray and gamma-ray astronomy
\cite{paper1,paper2,gorenstein2003,gorenstein2004}. We here
concentrate on phase Fresnel lenses (PFLs) using the terminology
of Miyamoto \cite{miyamoto}. In PFLs the optical thickness profile
is such that the phase shift is everywhere exactly the same,
modulo $2\pi$, as that of an ideal refractive lens. This leads to
a structure with the same periodicity as a zone plate but with
much higher efficiency.

PFLs are attractive for high energy astronomy, for which they
offer the prospect of extremely high angular resolution.
Performance can be diffraction limited, even in the gamma-ray part
of the spectrum. For diameters of a few metres this limit
corresponds to $\sim$ 0.1--1  micro arc second (\muas ), opening
up the possibility of imaging the black holes in distant galaxies,
for example. Equally important is the ability of PFLs to
concentrate flux from a large collecting area onto a small,
low-background, detector, resulting in orders of magnitude
improvement in sensitivity compared with current techniques.

 In microlithography and microscopy, PFLs and zone plates can provide
spatial resolution somewhat finer than the finest scale of the
structure of the lens, which can be\cite{eg2} of the order of a few tens of
nm.

PFLs  suffer from two major disadvantages. First, the focal
lengths tend to be long -- values of as much as several million
kilometers have been discussed for astronomy \cite{paper1}.
Second, in common with other diffractive optics, they are highly
chromatic. For PFLs, as for simple zone plates, the focal length
is inversely proportional to wavelength, so the good imaging
performance is only available over a narrow spectral band.

Surprisingly, with the developments of techniques for formation
flying of two spacecraft the first problem -- that of focal length
-- does not seem to be unsurmountable \cite{imdc-study-report}.
Furthermore accepting and using very long focal lengths
ameliorates chromatic effects \cite{spie2003}. There are, though,
practical limits to how far one can go in this direction. For
example the distance between spacecraft flying in formation and
carrying the lens and detector of an astronomical telescope is
limited by the thrust and fuel consumption of the thrusters that
are necessary to overcome gravity gradient effects and also by the
size of the detector necessary for a given field of view.

  \begin{figure}[h]\centerline{\scalebox{1}{\includegraphics{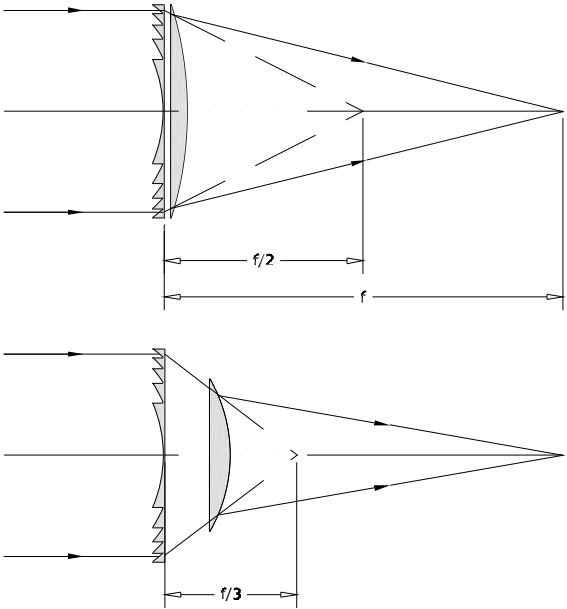}}} 
 \caption{ Use of a refractive lens to compensate for the chromatic
effects in a PFL. Upper : first order correction. Lower : second
order correction.
}
 \label{achrom_geom}
  \end{figure}

In some circumstances another consideration that can help overcome
the limitations associated with chromaticity is that
a PFL can be used at wavelengths over a wide band if the detector
distance is adjusted for each wavelength in turn. There is little loss
in image quality or efficiency \cite{paper1}, but of course the
observing time for any one wavelength is only a fraction of the
total time available.

It has long been known \cite{faklis-morris} that full achromatic
correction of the dispersion of a diffractive component is
possible using another diffractive component in  the configuration
originally proposed by Schupmann \cite{schupmann}. However this
arrangement is unlikely to be useful in the X/gamma-ray case
because it requires a further optical component to image the
aperture of one diffractive component onto the other. This
additional optical component can be smaller, but it has itself to
be achromatic. Bennett \cite{bennett76} has discussed the fact
that there are intrinsic limits on correction of diffractive
elements by other diffractive elements.

 \begin{figure}[h]\centerline{\scalebox{0.75}{\includegraphics {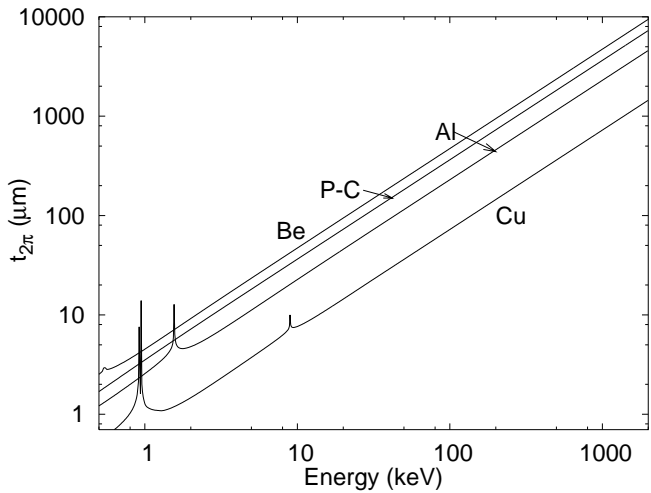}}} 
 \caption{  The thickness,  $t_{2\pi}$,  necessary to
produce a phase shift of 2$\pi$ for different materials (P-C
indicates Polycarbonate).
}
 \label{t2pifig}
  \end{figure}

Of interest here us the possibility, pointed out in a number of
papers \cite{paper2,wang, gorenstein2004} of making an achromatic
X-ray or gamma-ray system by combining a diffractive component with
a refractive one. The advantages and limitations of these schemes
are examined here. Emphasis is placed on astronomical
telescope designs, but reference is also made to laboratory
applications.

 \begin{figure}[h]\centerline{\scalebox{0.75}{\includegraphics {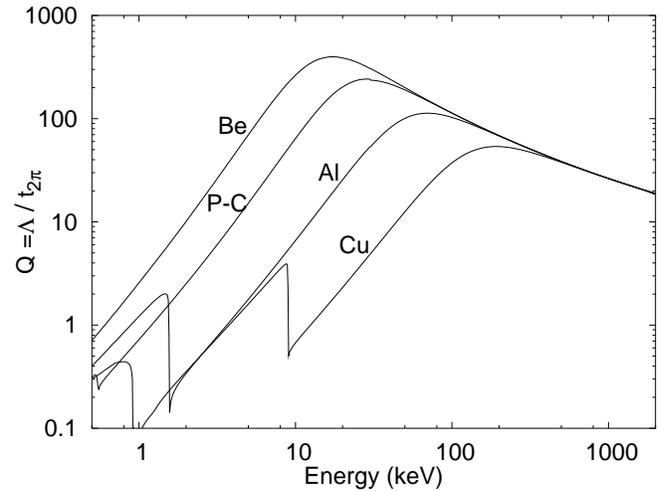}}} 
 \caption{The absorption length divided by  $t_{2\pi}$ for the materials in Fig. \ref{t2pifig}.
 }
 \label{qfig}
 \end{figure}

\section{ Diffractive/refractive achromatic designs}
\label{design-section}

 At x-ray and gamma-ray wavelengths the real part
of the refractive index of a material is usually written $\mu=1-\delta$, where $\delta$ is
small and positive ($10^{-3}$--$10^{-10}$), corresponding to refractive indices slightly
less than unity. At high energies \cite{note} and away from absorption
edges, $\delta$ is approximately proportional to $\lambda^2$, leading to a variation of the
 focal length of a refractive lens $D=(\lambda/f_r)(df_r/d\lambda)\sim -2$.
Wang et al.\cite{wang}  have suggested the use of the very much larger anomalous dispersion
in the region of the K and L absorption edges, but this occurs only in very narrow bands.
We here concentrate here on what can be achieved with $D\sim -2$.

For a diffractive lens $D=-1$, so it follows \cite{paper2} that
one can form an achromatic pair with combined focal length $f$ can
by placing a diffractive component in contact with a refractive
one (Fig. 1a). If, as in Fig. 1b, the two components are separated
\cite{gorenstein2004}  by a distance $S$, then $d^2f/d\lambda^2$
can be made zero as well as $df/d\lambda$ ~($f$ is the focal
length of the combination measured from one of the components;
here we use the first). However we shall see in section
\ref{design-params} that some degree of transverse chromatic
aberration is inevitable if the separation is non-zero.

 The problem with either of these forms of refractive correction
 is that in most practical circumstances the
refractive lens must be very thick. Suppose that the diffractive
component is operating in order $N_d$ (so that for a PFL the steps
correspond to a phase shift $N_d 2\pi $; usually $N_d=1$) and that
it has $N_z$ annuli (or zone pairs for a zone plate). It will
introduce a differential phase shift $N_zN_d 2\pi$ between the
centre and the periphery. In both the  contact and separated pair
cases, the refractive component must counteract half of this
difference. Thus it must have a central thickness $\mt2pi
N_zN_d/2$, where \t2pi is the thickness which introduces a phase
shift of $2\pi$.

This thickness should be compared with the attenuation length, $\:\Lambda$, of the
material.  Note that the relevant value of $\Lambda$
is one which includes the effects of losses by incoherent
(Compton)
 scattering as well as by photoelectric absorption and,
 for high energies, pair production. Fig.~2 shows \t2pi \
for various materials and in Fig. 3 the ratio $Q=\Lambda/\mt2pi $
is plotted. If $N_d N_z  \stackrel{>}{\sim} Q$
%
then correction by a simple refractive lens as in Fig.~1 would
require a lens in which the absorption losses at the centre would
be large. Consequently there is a practical limit on the size of a
diffractive lens that can be corrected by a simple refractive
element; its diameter, $d$, must be less than about $k(f_0
\lambda_0 Q)^{1/2}$. Here $k$ is $2\surd 2$ for the contact pair
case and $4/\surd 3$ for a separated pair and the subscript zero
refers to values at the nominal design energy. It can be shown
that the limit used here corresponds to requiring a mean
transmission of $1-e^{-1}$, or 61\%.

$Q$ is effectively the largest number of zone (pairs) in a
diffractive lens which can be corrected in this way (assuming
$N_d=1$). As shown in Fig.~3, $Q$ can reach a few hundred, the
highest values being attained with materials of low atomic number
($Z$) and with photons of energy approximately $10-100$ keV.
Within this energy band, and provided that a large focal ratio
$f/d$ requiring a limited number of zones is acceptable, the
simple configurations shown in Fig. 1 can be used. The design and
performance of such systems are discussed in the Subsection 4A.

Where the absorption of a simple refractive corrector would be
prohibitive, the natural solution is to reduce the thickness in
steps. The performance of such configurations is examined in
Subsections 4B and 4C.

\section{Design parameters}
\label{design-params}

 For an achromatic  contact doublet with net
focal length $f$ which combines a diffractive component, for which
$D=-1$, with a refractive one having $D=-2$,  the focal lengths of
the two components need to be $f_d=f/2$ and  $f_r=-f$
respectively.

If the two lens components are not in contact but separated by
distance $S$, the transfer matrix between the input aperture and a
detector plane distance, $f_0$ beyond it, is
\begin{eqnarray}
\lefteqn {\left(\begin{array}{cc} A & B \\ C & D \end{array}\right),}  \nonumber
\end{eqnarray}
 where, assuming $D=-2$,
\begin{equation}
\left.
{ \begin{array}{ccl}
A & =  & 1- {\displaystyle{{{\mathcal{E} f_0 f_1   + \mathcal{E}^2 f_0 f_2 - f_0 S - \mathcal{E} f_1 S + S^2} \over {\mathcal{E}^3 f_1 f_2}} }} \\
B & =  &  f_0 - S\ {\displaystyle{ \frac{f_0 - S}{\mathcal{E}^2 f_2}  }} \\
C & =  & {\displaystyle{ - \: {{\mathcal{E} f_1 + \mathcal{E}^2 f_2 - S}\over {\mathcal{E}^3f_1 f_2} }}} \\
D & =  &  1-    {\displaystyle{ {S\over{\mathcal{E}^2 f_2}} }}
\end{array} } \right\} ,
\end{equation}
and where $\mathcal{E}$ is the photon energy normalised such that $\mathcal{E}=1$ for $\lambda=\lambda_0$.
For parallel incident radiation  to be focussed at $f_0$, we set
the term $A$ to zero. Requiring that $df/d\mathcal{E}=d^2f/d\mathcal{E}^2=0$, as discussed
above,
leads to the configuration shown in Fig.~1b in which
\begin{equation}\label{gorenstein_params} \left.
{\begin{array}{rcl} S & = & f_0/9 \\ f_d & = & f_0/3 \\ f_r & = &
-(8/27)\, f_0
\end{array}  } \right\} .
\end{equation}
However the term $B$, which dictates the plate scale, is then
$f_0(1+1/(3\mathcal{E}^2))$. The energy dependence of $B$ implies
that there is transverse chromatic aberration (lateral colour). If
one places the alternative constraints $A=
df_0/d\mathcal{E}=dB/d\mathcal{E}=0$, the contact pair ($S=0$)
case discussed above is the only solution.

The transverse chromatic aberration is not necessarily an
over-riding problem. It has little impact if the lens is used
primarily as a flux collector -- one of the important potential
astronomical applications. Moreover, in imaging applications in which
 the detector has the capability
of determining the energy of each photon with adequate resolution,
the effect can be corrected during data analysis. However, it may
limit the usefulness of this configuration in some applications,
for example, in microlithography.


\begin{table}[h]
{\bf \caption{Example of a lens using unstepped refractive
correcting elements for an astronomical application. Nominal
energy 78.4 keV; focal length $1\times 10^8$m.
 \label{examples-non-stepped-table}}}
\begin{center}
\begin{tabular}{lccc}\hline
Configuration        & Singlet&Contact&Separated\\
\hline
                                 &      &          &               \\
Diffractive component:-          &      &          &               \\
 \hfill Diameter (mm)            & 1310 &  1310    &   1070        \\
 \hfill Focal length (m)             & $10\times 10^7$   & $5\times 10^7$ & $3.33\times 10^7 $ \\
 \hfill  Pitch (min)  ($\mu $m)   & 2420 &  1210    &     985       \\
 Separation (m)                   &      &  0       &  {$1.11\times 10^7$}    \\
Refractive component:-           &      &          &               \\
 \hfill  Diameter (mm)           &      & 1310     &     715       \\
\hfill   Focal length (m)        &      & $-10.0\times 10^7$ & $-2.96\times 10^7 $\\
 \hfill  Thickness (mm)          &      &  39      &      39       \\
Bandpass (keV)                   & 0.1  &     4.7 & 27.7           \\
  \hline
\end{tabular}
\end{center}
The refractive component is assumed to be made of beryllium.   The
diameters of the achromatic combinations are chosen according to
the guidelines in the text, and so the theoretical efficiency is
61\% in each case. The singlet is assumed to have the same
diameter as the contact pair, though absorption does not preclude
it being made larger. The bandpass quoted is full width at half
maximum of the on-axis response.
\end{table}


\begin{table}[h]
{\bf \caption{Example of a lens using unstepped refractive
correcting elements corresponding to that in Table
\ref{examples-non-stepped-table} but for a laboratory application.
Nominal energy 10 keV; focal length 1 m. Other details as
Table~\ref{examples-non-stepped-table}.
 \label{examples-non-stepped-table_2}}}
\begin{center}
\begin{tabular}{lccc}\hline
Configuration        &Singlet&Contact&Separated\\
 \hline
                        &          &            &                   \\
Diffractive component:-          &            &                   \\
 \hfill Diameter (mm)    & 0.511    &  ~~~0.511  &    ~~0.417        \\
 \hfill Focal length     &1.0   &  ~~0.500       &  ~~0.333          \\
 \hfill  Pitch (min)  ($\mu $m) & 0.486      &  ~~~0.243    &  ~0.198      \\
 Separation (m)          &          &   ~0~~~~   &  ~~0.111          \\
Refractive component:-   &          &            &                   \\
 \hfill  Diameter (mm)   &          &~~~ 0.511   &  ~~0.278          \\
\hfill   Focal length (m) &          & $-1.0$     & $-0.296$          \\
 \hfill  Thickness (mm)  &          &   9.57     &     9.57          \\
Bandpass (keV)           & 0.03    &   ~~0.8~~~ &  ~~2.9~~~         \\
  \hline
\end{tabular}
\end{center}
\end{table}

\section {Imaging performance}

\begin{figure}[h]\centerline{\scalebox{0.75}{\includegraphics{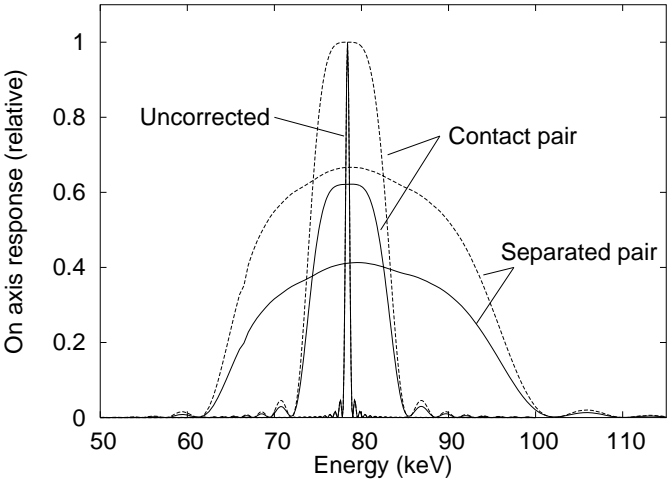}}} 
\caption{ The on-axis response of lenses for a nominal design
energy of 78.4 keV and with the parameters given in
Table~\ref{examples-non-stepped-table}. The dotted lines show the
response in the absence of absorption. Curves are  normalised so
the the peaks values are proportional to the effective areas. The
peak response of the separated pair is lower than that of the
contact pair because the absorption considerations discussed in
section 2 require a smaller diameter. Variations in image
brightness linked to changes of the diffraction limited resolution
with energy have been removed.  \newline \it{File:
skinner-achrom-fig4.eps} } \label{examples-non-stepped-figure}
 \end{figure}

The complex amplitude of the response at distance $f_0$ and radial
offset $y$ to parallel radiation falling at normal incidence on a
thin lens of arbitrary circularly symmetric profile,  is
\begin{eqnarray}
\int_{0}^{r^{2}}J_0 \left({{y\surd s}\over{\lambda f_0}}\right)
exp\left[ -{t(s)\over{\Lambda}}+i\pi \left({t(s)\over{\mt2pi
(\lambda)}} - {{s+y^2}\over{f_0\lambda}}\right)\right] ds. \nonumber
\end{eqnarray}
Here the total thickness $t(s)$ of the lens combination, is
specified in terms of the parameter $s=r^2$ rather than radius
$r$.
Assuming  the wavelength dependence of \t2pi \  corresponding to
$D=-2$, rewriting in terms of energy, $\mathcal{E}$, and allowing
for the possibility that the  incoming radiation is not parallel,
so that the phase $\phi$ will be a function of $s$, yields for the
term in square brackets
\begin{equation}
\left[-{t(s)\over{\Lambda}}+i\pi \left({t(s)\over{\mt2pi (\lambda_0)}}{1\over{\mathcal{E}}}-
{{s+y^2}\over{f_0\lambda_0}}{\mathcal{E}}+\phi (s)\right)\right].
\end{equation}

Numerical integration of this function provides simple and precise
evaluation of the response of a contact pair and, applied twice,
the response of a separated pair to on-axis radiation.

\subsection {Unstepped correctors }
\label{unstepped_subsection}

To illustrate the advantages gained with a simple unstepped correction lens, either in
 a contact pair and as part of a
separated pair, Table \ref{examples-non-stepped-table} presents
some example values for a space instrument having a very long focal length ($10^5$
km) and working at the astronomically interesting energy of 78.4
kev ($^{44}$Ti decay). Fig. \ref{examples-non-stepped-figure} shows
the on-axis response as a function of energy. The 0.1 keV bandpass of the uncorrected
lens can be increased to 4.7 keV (contact pair) or 27.7 keV (separated pair).
The bandwidth of the separated pair is sufficiently large that a single configuration
could observe simultaneously at the energy of the 78.4 keV $^{44}$Ti line
and at that of the 67.9 keV line which accompanies it, as well as covering a
large swath of continuum emission.

The on-axis point source response function is very close to the
Airy function of an ideal lens, but is modified slightly by the apodizing
effect of the radially varying transmission. The width of the point source
response function is reduced by a few percent and the intensity of the first
sidelobe is increased by $\sim$50\%.

The contact pair has extremely low geometrical
aberrations over a very wide field of view, as would be expected for a
thin lens with a very large focal ratio. For a separated pair, the
field of view is limited by vignetting at the corrector and this diameter
has to be made somewhat larger than the minimum 2/3 $d_d$ needed for on-axis radiation,
with a consequent increase thickness (and hence absorption). Within the vignetting-limited
field of view geometrical abberations are entirely negligible.

It is natural to ask whether the configurations considered here for space astronomy
 could be scaled to be useful for laboratory applications. Table \ref{examples-non-stepped-table}
show parameters for a 1 m focal length system working at 10 keV.
Concentrating on the separated pair, the thickness/diameter ratio
of the corrector lens would be 9.57 mm/0.278 mm = $\sim$34 in
place of 39 mm/715 mm = 0.055.  Mechanically this is very far from
a `thin lens' ! Because the radiation is always nearly paraxial,
optically the thin lens approximation used in the above analysis
remains a good first approximation, though no more than that. A
small fourth order term in the profile, amounting to 0.1\% of the
thickness, is sufficient to correct the on-axis response. Off-axis
aberrations then remain smaller than the diffraction-blurring over
a field 1.6 mm in diameter containing $>10^7$ resolvable pixels.
(In order to preserve the analogy with the astronomical case,
focussing of a parallel input beam is assumed here. The 50\%
encircled power diameter is used as a measure of resolution and it
has been assumed that the corrector is oversized by 10\% to reduce
vignetting.)

The examples quoted here represent approximately the extremes of size and of
high and low energy for which unstepped configurations are likely to be attractive.

  \begin{figure}[h]\centerline{\scalebox{0.4}{\includegraphics{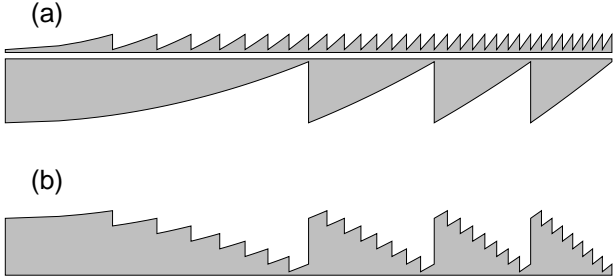}}} 

 \caption{ Profile of diffractive lens and stepped compensating refractive component.
 The example shown corresponds to $N_d=1,\: N_r=4$. (a) Two distinct
 components in contact. (b) A single component with the same thickness
 profile.
 }
 \label{profile}
  \end{figure}


  \begin{figure}[h]\centerline{\scalebox{0.4}{\includegraphics{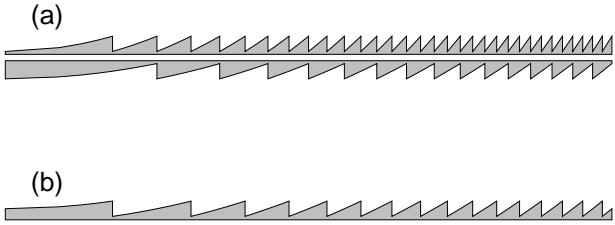}}} 

 \caption{ Same as Fig. \ref{profile}, but for $N_d=1,\: N_r=1$. The
 combined profile (b) is identical to that of a
 simple PFL, so clearly in this limit no compensation is achieved.\newline \it{File: skinner-achrom-fig6.eps} }
 \label{profile2}
  \end{figure}

\subsection {Stepped contact pair }

\label{stepped_subsection}

As has been suggested above, the natural solution to excessive
absorption in the refractive component is to reduce the thickness
of the refractive component in a stepwise fashion (Fig.
\ref{profile}). Ideally the steps would be an integer number,
$N_r$, times  \t2pi  \ but this cannot be exactly true for all
energies. We consider first contact pairs.

There are limits to the stepping process as can be seen by noting that in
the case of a contact pair in which the two components are made
from the same material, all that is really important is the total
thickness, $t(s)$. With $N_r = N_d =1$ the resulting thickness profile
of a contact pair would be as shown in Fig. \ref{profile2}, which
is that of a simple (chromatic) PFL of focal length $f$. At the
opposite extreme, for $N_r>N_zN_d/2$, no steps occur.

We will here consider the performance of a thin contact achromatic pair (Fig. 1a)
with intermediate degrees of `stepping', $N_d<N_r<N_zN_d/2$.

To illustrate this case we take the case of a lens for
imaging 500 keV gamma-rays  having a diameter of 2.048~m
(511 keV or the energies of other lines
of astrophysical interest could equally have been chosen).
 Figure~\ref{qfig} shows that  at this energy $Q$ is about 38 for
any material of which the atomic number is not too high. This
 independence of material arises when the dominant loss mechanism
is Compton scattering -- $\Lambda$ then depends
simply on the density of electrons, as does $\delta$. Thus although
the lens is here assumed to be made from polycarbonate, similar performance
could be obtained from many other materials. According to the
 criterion of section \ref{design-section} the maximum
 diameter for a simple unstepped corrector would be 0.78 m for a focal length of
 $8\times 10^8$ m.

  \begin{figure}[h]\centerline{\scalebox{0.75}{\includegraphics{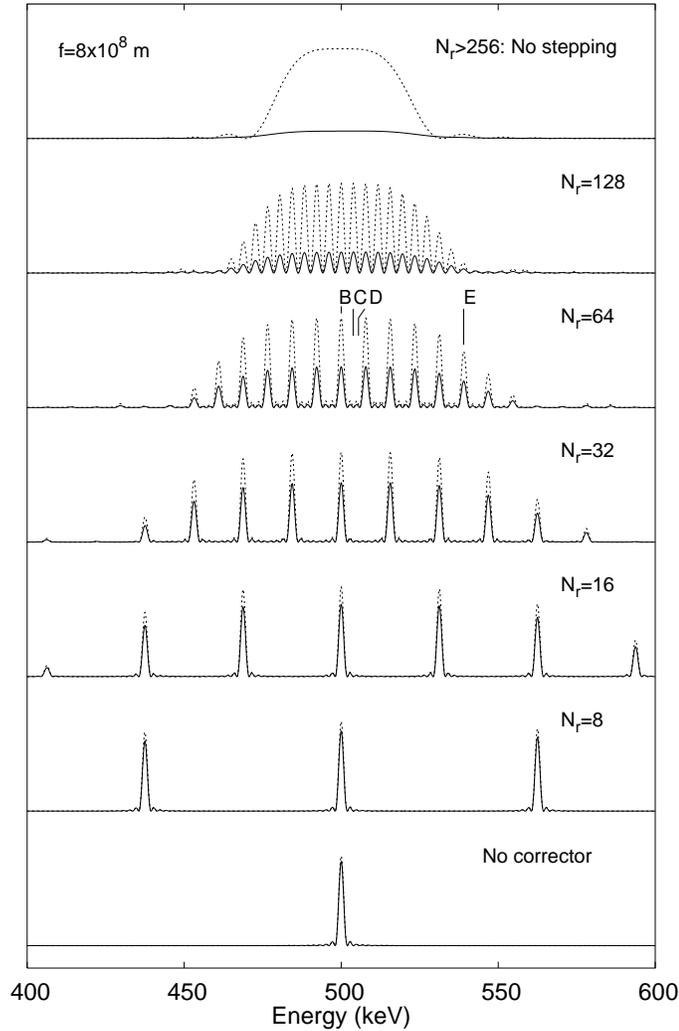}}} 
 \caption{ The on-axis response of achromatic contact pairs  with different degrees of stepping of the refractive
component, characterised by $N_r$, where the thickness is reduced
modulo $N_r t_{2\pi}$. The focal length and diameter are kept
constant. See Table 2 for detailed parameter values. The dashed
lines show the response if there were no absorption. It can be
seen that large values of $N_r$ give a  response which is more
nearly continuous as a function of energy but with absorption
losses which become more and more serious. The energies indicated
by B,C,D,E are those for which the point source response functions
are shown in Fig. \ref{fig_psf}.
}
\label{500keVlens-figure_1}
  \end{figure}

To illustrate the effect of
different design options we consider variations on a `reference' design
with $N_r=64$ and a focal length of  $8\times 10^8$ m.
The assumed parameters are given in Table \ref{500keVlens-table}.

\begin{figure}[h]\centerline{\scalebox{0.75}{\includegraphics{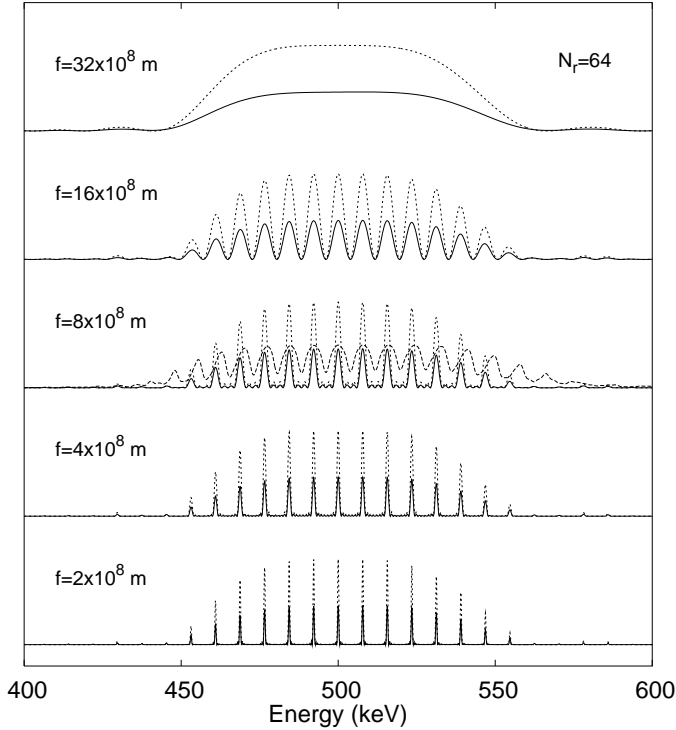}}} 
 \caption{ As Fig.~\ref{500keVlens-figure_1} but showing the effect of differing focal lengths for a
 particular  value of $N_r$ (=64).
 Shorter focal lengths require a more powerful correction lens which,
 for a  given $N_r$, is
 divided into a larger number of zones, resulting in narrower individual peaks in the energy response.
 In the case of the central plot showing the reference design,
 which would otherwise be a repeat of the corresponding plot in  Fig.~\ref{500keVlens-figure_1},
 the dashed line indicates how
 the response differs when seen by a detector integrating over a 5 mm radius.
 }
\label{500keVlens-figure_2}
  \end{figure}

The calculated on-axis response for different values of $N_r$, either side of the
reference design value, is shown in Fig.~\ref{500keVlens-figure_1}
and the effect of different focal lengths
in Fig.~\ref{500keVlens-figure_2}.
A series of peaks occurs within an envelope which is dictated by
the uncorrected higher order terms in the Taylor expansion of $f$.
The total response, integrated over all the peaks
is shown in Fig. 9. As would be expected, for large $N_r$ the absorption
negates the gain from having more peaks, the optimum
in the example case  being at $N_r\sim50$ according to this criterion.

The spacing of the peaks is $E_0/N_r$, where
$E_0$ is the nominal design energy.
Their fractional width is inversely proportional to $N_z$, the number of annuli in the
refractive component, just as the finesse of a Fabry-Perot
interferometer depends on the number of
interfering beams. Thus if $N_r$ (and hence the thickness and the absorption loss)
is kept constant, long focal lengths widen the peaks and increase the integrated response.

 The point source response function at
energies between the peaks of the on-axis response is obviously an
important consideration. For energies away from these peaks the
flux is spread into an extended diffraction pattern with
concentric rings having a characteristic width which is comparable
in scale with the width of the ideal PSF (Fig.~\ref{fig_psf}).

For practical applications in which an image is to be acquired,
one can consider two possible situations. If one has an
energy-resolving detector capable of recording separate images for
each of the energies at which the image response differs
significantly, then post processing can recover most of the
imaging performance of an ideal lens. If however the energy
resolution of the detector is insufficient to isolate the peaks in
the on-axis response, then the effective angular resolution will
be degraded by mixing of the different responses. The point source
response function will then be one which is averaged over all
energies within the band, as shown by the continuous curves in
Fig.~\ref{fig_psf}. In the example used here, the half power
diameter would be degraded from 0.27\muas (about 1 mm) to
1.8\muas.


\begin{table}[h]
{\bf \caption{Width of point source response function of the
lenses considered in Table~\ref{500keVlens-table} and in
Fig.~\ref{500keVlens-figure_1} }}\begin{center}
\begin{tabular}{lcc}\hline
       &Full Width&Half Power\\
  Optical system      &Half Maximum& diameter  \\
   and energy     &(micro arc secs)&(micro arc secs)\\
\hline
Ideal diffraction-\\
limited lens     &     0.265    &    0.270     \\
Achromat with \\
absorption (500 keV) &     0.260   &     0.275     \\
Ditto (averaged \\
response 450--550 keV)  &    0.265     &    1.80  \\
 \\ \hline
\end{tabular}

\end{center}
\end{table}



\begin{table}[h]
{\bf \caption{Parameters for the example gamma-ray lenses for
astronomical applications for which the performance is shown in
Fig.~7--9. Parameters in common are:- nominal energy: 500 keV,
diameter: 2048 mm, material: polycarbonate. The `reference' design
values are in boldface.
 \label{500keVlens-table}}}
\begin{center}
\begin{tabular}{cccccc}\hline
Figs. &Focal       & Diffractor  & \multicolumn{3}{c}{Refractor}      \\
      & length, $f_0$          & Zones       &  Step          & Zones        & Thickness \\
       &    (m)         &       $N_z$ &       $N_r$     &       $N_{zr}$   &  (mm)      \\
\hline
      &                 &   \hfill \vline &      256        &      1         &  605 \\
     &                  &   \hfill \vline &      128        &     2          &  302 \\
     &                  &   \hfill \vline &     \bf{64}     &   \bf{4}      &  \bf{151} \\
  7,9&{$8\times 10^8$}  &\hfill 512\hfill \vline &     32   &     8          &   75 \\
     &                  &   \hfill \vline &       16        &    16          &   38 \\
     &                  &   \hfill \vline &        8        &    32          &   19 \\
     &                  &   \hfill \vline &    \multicolumn{2}{c}{(none)~ ~} & 2.4 \\
\hline
     &{$2\times 10^8$}  &\hfill 2048 \hfill \vline &               &              &    \\
     &{$4\times 10^8$}  &\hfill 1024 \hfill \vline &               &              &    \\
   8 &$\bf{8\times 10^8}$  &\hfill \bf{512} \hfill \vline &   \bf{64}       & \bf{4}    &\bf{151}\\
     &{$1.6\times 10^9$}  &\hfill 256 \hfill \vline &               &               &    \\
     &{$3.2\times 10^9$}  &\hfill 128 \hfill \vline &               &                &    \\
  \hline
\end{tabular}
\end{center}

\end{table}

If the compound lens is to be used to increase the
bandpass of a projected image, in microlithography for example, a
degradation of resolution like that in Fig.~\ref{fig_psf} will usually have to
be accepted. It is possible  that the multiple peaks
in the  response might be adjusted to correspond to
multiple lines in a fluorescent X-ray source or in synchrotron radiation
from an undulator, or to multiple astrophysical lines such as the 57.9/78.3 keV
Ti line pair discussed in section 4A.

Where the compound lens is used as a flux concentrator, a small detector
will receive the multi-peaked response seen in Fig.~\ref{500keVlens-figure_1}, while
for a larger one the troughs will be filled in.

  \begin{figure}[h]\centerline{\scalebox{0.75}{\includegraphics{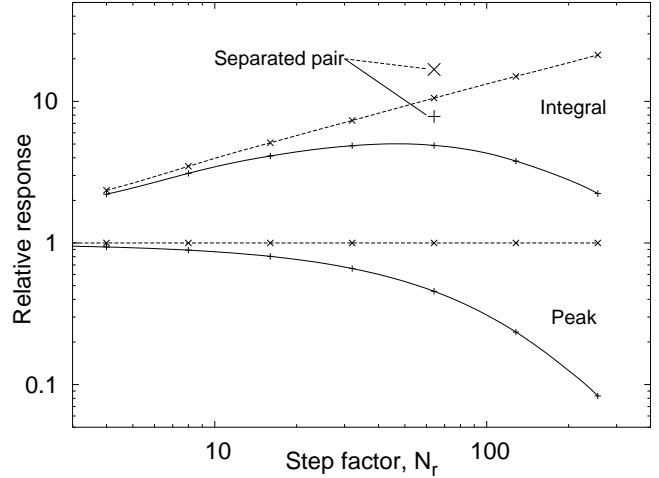}}} 
 \caption{ The total on-axis response to a point source, integrated over energy, for the lens in Fig.
 \ref{500keVlens-figure_1} and the corresponding peak response, shown for different degrees
 of stepping. The dashed lines again show the results in the absence of absorption. Points
 for the separated pair described in section \ref{separated_pair_section} are also shown.
  }
 \label{achrom_resp}
  \end{figure}

\subsection{Stepped separated pair}

\label{separated_pair_section}

Stepping can, of course, be used with a separated achromatic pair.
We content ourselves here with presenting (Fig.~\ref{fig_seppair}) the on-axis
response of a separated lens pair version of the reference gamma-ray lens
design considered in section \ref{stepped_subsection}.

As for a contact pair, the response as a function of energy has multiple peaks
whose number and width are a function of the degree of stepping. Again
between these peaks the flux is spread over a larger region, so the PSF is again
broadened unless photons with these intermediate energies can be ignored.
The detailed processes which give rise to the peaks and troughs are different because
there is no longer a one-to-one correspondence between regions of the two optical
elements. Although  in the region of ${E}_0$ the spacing of the peaks is
the same as for the contact pair, the spacing is not constant but varies as $E^2$.
Their width is typically a factor of $\sim 1.5$ smaller than for the reference contact pair
case, which has the same $N_r$ and the same absorption losses. In consequence
the gain in integrated response is only a factor of 1.65.

  \begin{figure}[h]\centerline{\scalebox{0.75}{\includegraphics{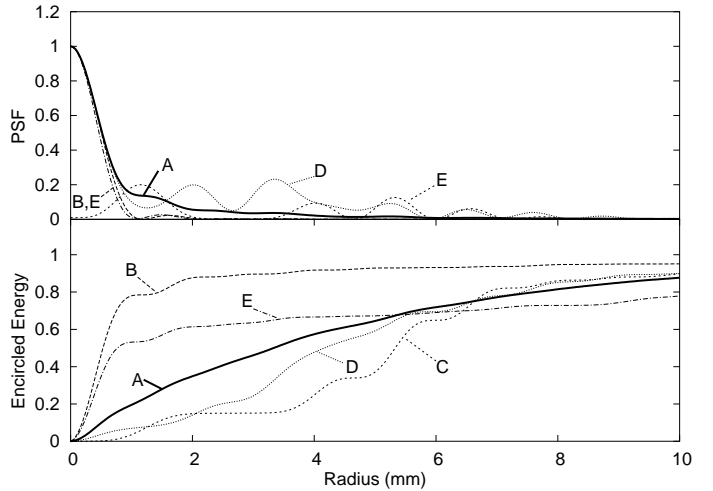}}} 
 \caption{ (a) Point source response function of the lens in Fig. \ref{500keVlens-figure_1},
 averaged over energies 450-550 keV  (A, bold line) and at selected energies (B: 500.0 keV,
 C: 503.8 keV, D: 505.1, E: 539.0 keV). Data have been normalised to the central peak
  except for 503.8 keV where there is none. (b) The corresponding  fractional encircled
 power as a function of radius. Absorption is taken into account.
  }
 \label{fig_psf}
  \end{figure}

\section{Practicability}

In the preceding sections  various designs and possible materials
have been mentioned without consideration of the practicability of
fabricating the necessary  structures. Although extensive
experience exists with the fabrication of X-ray lenses, so far
such lenses have been {\it either} diffractive {\it or}
refractive, and have been much smaller than those proposed here
for astronomical applications.

Issues of absorption are not usually critical for diffractive
lenses  (either Fresnel lenses or the simpler Phase Zone Plates in
which only two discrete thicknesses are employed). Fig.~2 shows
that above a few keV many materials will have absorption which is
not too important for the maximum thickness required (\t2pi \  or
$\mt2pi /2$ respectively). Silicon has often been the material of
choice because of the sophisticated micromachining techniques
which are available for this material (e.g.
\cite{silens}
 ), but high density
materials are sometimes preferred because the reduced thickness
necessary allows finer scale structures and hence higher spatial
resolution for microscopy applications (e.g.
\cite{eg1}
). Phase Zone Plates with gold structures with 1.6
$\mu$m thick, corresponding to $\mt2pi /2$ \ at 9 keV, and with
finest periods of 0.2$\mu$m or less are advertised commercially
\cite{xradia}.

For refractive lenses, now widely used following the work of
Snigirev et al. \cite{snigirev},  absorption is more important and
low atomic number materials are preferred. Beryllium
\cite{belens}, and even Lithium \cite{lilens}, have been employed,
though plastics also seem to offer a good compromise. Again small
diameter lens stacks are commercially available \cite{adelphi}.

Deviations from the nominal profile will degrade efficiency.
Nevertheless efficiencies of 85\% of the theoretical value have
been reported for a multilevel nickel lens at 7 keV \cite{eg1} and
$>$75\% of theoretical efficiency is apparently obtainable for
Phase Zone Plates\cite{xradia}. For structures having a larger
minimum pitch such as those discussed here, close tolerances and
near ideal profiles should be easier to achieve.

Where the refractive component has a thickness of many times \t2pi
\ (large $N_r$), the uniformity of the material should also be
considered. SAXS (small angle X-ray scattering) measurements
suggest that certain materials which are structured but
non-crystalline should be avoided and point to amorphous materials
of which Polycarbonate is a good example \cite{saxs}. For
materials such as glass and plastics, experience in manufacturing
optical components for visible light provides evidence that
inhomogeneities (or at least those on a scale larger than the
wavelength of visible light) present no overriding problem and
also demonstrates the feasibility of attaining the precision
necessary for high energy lenses made from these materials. We
note that \t2pi \ provides a measure of the precision required in
the different cases.  For typical materials used for transmission
optics \t2pi \ is of the order of 1 $\mu$m for visible light.
Fig.~1 shows that for X-rays and gamma-rays it is almost always
much larger and fabrication tolerances and demands on homogeneity
are correspondingly more relaxed. For example, at the highest
energies considered here ($\sim$ 1 MeV), achieving $\lambda/40$
accuracy demands no more precision than a few tens of microns
compared with $\sim 0.02 \mu$m in the visible!

Finally, in connection with the construction of large scale lenses
one can note the development of thin etched glass membrane
diffractive lenses of many square meters \cite{eyeglass}.

\section{Conclusions}

The various achromatic combinations discussed here have significant advantages
over simple PFLs. Absorption limits the size of lens which can be corrected. It is
least problematic for energies in the range $\sim$10--100 keV and for long focal lengths.
Stepping helps reduce the focal length for which a given diameter lens
can be corrected, but works best at a comb of energies within the bandpass.
If the  angular resolution is not to be degraded it is
necessary to employ a  detector with energy resolution good enough that the flux at energies
between the peaks in the response can be ignored or treated separately.

The separated pair solution has considerably wider response with
similar performance in other respects. However for astronomical
applications it has the major disadvantage of requiring the
precise (relative) positioning of three satellites instead of just
two.

\section*{Acknowledgements and Notes}
The author is very grateful to P.~Gorenstein and to L.~Koechlin for helpful
discussions.

The mention of particular suppliers in this paper is for
information only and does not constitute an endorsement of their
products by the author or his institution.

  \begin{figure*}[h]\centerline{\scalebox{0.7}{\includegraphics{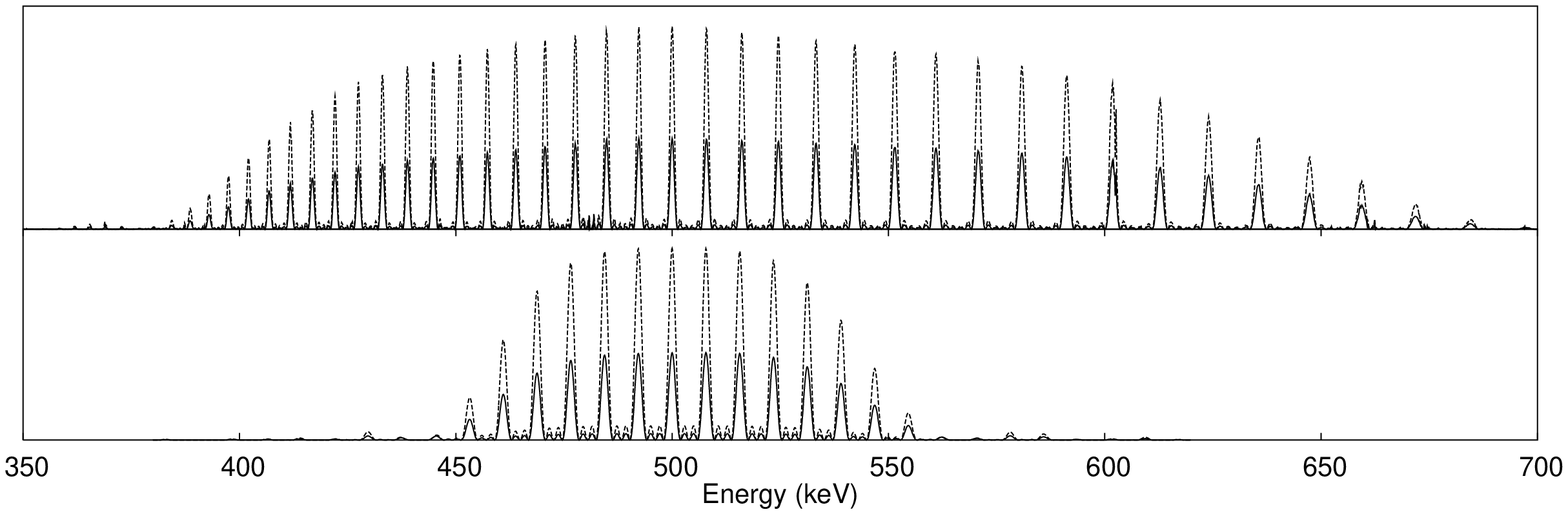}}} 
 \caption{ The on-axis response of a separated achromatic pair (above) compared with that of a contact pair.
 Parameters are given in Table~2. \newline
}
 \label{fig_seppair}
  \end{figure*}

\begin{thebibliography}{}
%
%
\bibitem{eg1} 
E. Di Fabrizio, F. Romanato, M. Gentill, S. Cabrini, B. Kaulich,
J. Susini, and R. Barrett,
 ``High-efficiency multilevel zone plates for keV X-rays'' 
Nature {\bf 401,} 895--898 (1999).
%
\bibitem{eg2}
W. Yun, B. Lai, A.A. Krasnoperova, E. Di Fabrizio, Z. Cai, F.
Cerrina, Z. Chen, M. Gentilli, and E. Gluskin, ``Development of
zone plates with a blazed profile for hard x-ray applications''
Rev. Sci. Instrum. {\bf 70,} 3537--3541 (1999).
\bibitem{eg3}   
E.H. Anderson, D.L. Olynick, B. Harteneck, E. Veklerov, G.
Denbeaux, W. Chao, A. Lucero, L. Johnson, and D. Attwood,
``Nanofabrication and diffractive optics for high resolution X-ray
applications'' J. Vac. Sci. Tech. B. {\bf 18,} 2970--2975 (2000).
%
\bibitem {paper1}
G. K. Skinner,  ``Diffractive-refractive optics for high energy
astronomy I Gamma-ray phase Fresnel lenses.'' Astron. \&
Astrophys. {\bf 375,} 691--700 (2001).
%
\bibitem {paper2}
G. K. Skinner, ``Diffractive-refractive optics for high energy
astronomy II Variations on the theme'' Astron. \& Astrophys. {\bf
383,} 352--359 (2002).
%
\bibitem {gorenstein2003}
P. Gorenstein, ``Concepts: X-ray telescopes with high-angular
resolution and high throughput'' in {\it X-ray and Gamma-Ray
Telescopes for Astronomy,} J. E. Truemper and H. D. Tananbaum,
eds., Proc. SPIE {\bf 4851,} 599--606 (2003).
%
\bibitem {gorenstein2004}
P. Gorenstein, ``Role of diffractive and refractive optics in
X-ray astronomy'' in {\it Optics for EUV, X-Ray, and Gamma-Ray
Astronomy,}  O. Citterio and S. L. O'Dell, eds., Proc. SPIE {\bf
5168,} 411--419 (2004).
%
\bibitem {miyamoto}
K. Miyamoto,  ``The Phase Fresnel lens'' J. Opt. Soc. Am., 51,
17--20 (1961).
%
\bibitem{imdc-study-report}
Integrated Mission Design Center Studies \textit{Fresnel Lens
Gamma-ray Mission} Jan 7--10, 2002, and  \textit{Fresnel Lens
Pathfinder} Jan 28--29, 2002, NASA Goddard Space Flight Center,
Greenbelt, MD.
%
\bibitem {spie2003}
G. K. Skinner,``Fresnel lenses for Xray and gammaray astronomy''
in {\it Optics for EUV, X-Ray, and Gamma-Ray Astronomy,}  O.
Citterio and S. L. O'Dell, eds., Proc. SPIE {\bf 5168,} 459-470
(2004).
%
\bibitem{faklis-morris} %
D. Faklis and G.M. Morris,``Broadband imaging with holographic
lenses", Opt. Eng. {\bf 28,} 592--598 (1989).
%
\bibitem{schupmann} %
L. Schupmann, \textit{Die Medial Fernrohre: Eine neue Konstrucktion fur
grosse astronomische Instrumente} (B. G. Tuebner, Leipzig, 1899).
%
\bibitem{bennett76} %
S. J. Bennett, ``Achromatic combinations of hologram optical
elements", Appl. Opt. {\bf 15,} 542--545 (1976).
%
\bibitem{wang}
Y. Wang, W. Yun, and C. Jacobsen,  ``Achromatic Fresnel optics for
wideband extreme-ultraviolet and X-ray imaging'' Nature {\bf 424,}
50-53 (2003).
%
\bibitem{note}
Mixing energy and wavelength notations seems inevitable here; the
former is more generally recognisable the X-ray and gamma-ray
domain, but the present work depends heavily on the wave-like
nature of the radiation.
%
\bibitem{silens}
B. N\"ohammer, J. Hoszowska, H.-P. Herzig and C. David,
``Zoneplates for hard X-rays with ultra-high diffraction
efficiencies" J. Phys. IV France  {\bf 104,} 193--196 (2003).
%
\bibitem{xradia}
Xradia Inc., 4075A Sprig Dr. Concord CA 94520 USDA.
http://www.xradia.com/zpl-pd.htm
%
\bibitem{snigirev}
A. Snigirev, V. Kohn, I. Snigireva and B. Lengeler, ``A compound
refractive lens for focusing high-energy X-rays 49" Nature, {\bf
384,} 49--51 (1996).
%
\bibitem{belens}
C.~G. Schroer,  M. {Kuhlmann}, B. {Lengeler},  T.~F. {G{\"
u}nzler}, O. {Kurapova},  B. {Benner},   C. {Rau}, A.~S.
{Simionovici},  A.~A. {Snigirev}, and I. {Snigireva}, ``{Beryllium
parabolic refractive x-ray lenses}'' ,  in {\it Microfabrication
of Novel X-Ray Optics,} D. C. Mancini,  ed., Proc. SPIE {\bf
4783,} 10--18 (2002).
%
\bibitem{lilens}
E. M. Dufresne, D. A. Arms  R. Clarke, N. R. Pereira, S. B.
Dierker and D. Foster, ``{Lithium metal for x-ray refractive
optics}'', Appl. Phys. Let., {\bf 79,} 4085--4087 (2001).
%
\bibitem{adelphi}
Adelphi Technology, Inc. 981-B Industrial Road San Carlos, CA
94070 USA. http://www.adelphitech.com/lenses.html
%
\bibitem{saxs}
B. Lengeler, J. T{\" u}mmler, A. Snigirev, I. Snigireva and C.
Raven ``{Transmission and gain of singly and doubly focusing
refractive x-ray lenses}'' J. Appl. Phys. {\bf 84,} 5855--5861
(1998).
%
\bibitem{eyeglass}
J. Early, R. Hyde, and R. Baron, ``Twenty meter space telescope
based on diffractive lens'' in {\it Optics for EUV, X-Ray, and
Gamma-Ray Astronomy,}  O. Citterio and S. L. O'Dell, eds., Proc.
SPIE {\bf 5168,} 459--470 (2004).


\end{thebibliography}
\end{document}